%% file: main.tex
\documentclass{article}
\usepackage{locata,amsmath,graphicx,url,times}

\input{mypreamble}

\title{Multiple source direction of arrival estimation using subspace pseudointensity vectors}

\name{Alastair H. Moore%
    \thanks{This work was supported by the Engineering and Physical Sciences Research Council {[}grant number EP/M026698/1{]}.}%
    \address{Imperial College London\\
			Department of Electrical and Electronic Engineering\\
			London, SW7 2AZ}%
}

\input{common_symbols.tex}

\input{symbol_definitions.tex}

\begin{document}

\newcommand{\acro}{\acrodef}\newcommand{\acroindefinite}{\acrodefindefinite}\input{myacronyms.txt}

\ninept
\maketitle

\begin{sloppy}

\begin{abstract}
The recently proposed subspace pseudointensity method for direction of arrival estimation is applied in the context of Tasks 1 and 2 of the \LOCATA\ Challenge using the Eigenmike recordings. %
Specific implementation details are described and results reported for the development dataset, for which the ground truth source directions are available. %
For both single and multiple source scenarios, the average absolute error angle is about \ang{9} degrees.

\end{abstract}

\begin{keywords}
direction of arrival estimation, spherical microphone array, multiple sources, array processing
\end{keywords}

\section{Introduction}
\label{sec:intro}

In recent years many \ac{DOA} estimation methods operating in the \ac{SH} domain have been proposed \cite{Rafaely2004,Teutsch2005,Khaykin2009,Jarrett2010,Rafaely2010,Sun2012,Evers2014,Nadiri2014,Moore2015a,Pavlidi2015,Moore2016b}. %
These methods transform the signals captured by a spherical microphone array into a representation in which the steering vectors are independent of the specific array geometry and of frequency. %
In practice, the size of the sphere, the number of microphones and their arrangement affect the \ac{SNR} as a function of frequency, limiting the usable frequency region. %

The \ac{SSPIV} method \cite{Moore2016b} uses \acl{FS} \cite{Khaykin2009,Nadiri2014}, in addition to conventional time-smoothing, to estimate the \ac{SH} covariance matrix of the sound field over a range of frequencies. %
This reduces the coherence caused by multipath progation and allows \ac{DOA} estimates to be obtained from shorter observation intervals, i.e. with less time-smearing, than those based on a single frequency bin.

Exploiting \ac{WDO} \cite{Rickard2002}, it has become common to estimate the \ac{DOA} of a single source in each \ac{TF} region before combining these estimates to estimate the number of sources and their directions. %
Some methods \cite{Nadiri2014,Moore2015a,Pavlidi2015} specifically test each \ac{TF} region for the validity of this assumption before deciding whether to include it in the overall estimation. %
However, this can lead to relatively few \ac{DOA} estimates being retained. %
In contrast the \ac{SSPIV} method includes all the local estimates on the assumption that, on average, even erroneous local estimates will tend to cluster around the true \acp{DOA}. %

The \ac{SSPIV} method is fully described in \cite{Moore2016b} and so will not be repeated in this work. %
Rather we focus on a block-level description of the overall system in Section~\ref{sec:system-overview} and specific implementation issues in Section~\ref{sec:implementation}. %
In Section~\ref{sec:results} results are reported for Tasks 1 and 2 of the \LOCATA\ Challenge development dataset \cite{Lollmann2018}. %
Finally, Section~\ref{sec:conclusions} concludes the paper.

\section{System description}
\label{sec:system-overview}
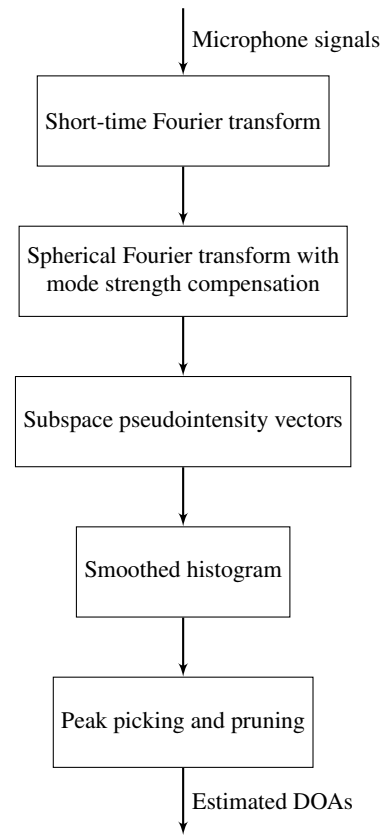
\begin{figure}[t]
  \centering
    \input{block_diagram}
  \caption{System diagram}
  \label{fig:system-diagram}
\end{figure}

Figure~\ref{fig:system-diagram} shows an overview of the system used to perform \ac{DOA} estimation for the \LOCATA\ Challenge. %
The 32-channel Eignemike signals are first transformed to the \ac{STFT} domain before further transformation to the \ac{SH} domain \cite{Rafaely2015}. %
Mode strengh compensation is used to account for the scattering effect of the rigid spherical geometry of the mirophone array. %
\acp{SSPIV} are calculated at each \ac{TF} bin according to \cite{Moore2016b}. %
i.e. at each \ac{TF} bin a Cartesian unit vector is obtained whose direction points towards an estimate of the \ac{DOA} for that \ac{TF} bin.

A 2-dimensional histogram of all the \acp{SSPIV}' directions is obtained in terms of the azimuth and inclination (defined as per the \LOCATA\ definition of elevation) components. %
The histogram grid is regularly spaced in azimuth and inclination which means that the equivalent surface area of each patch on the sphere is non-uniform. %
The raw histogram is smoothed with a Gaussian kernal, defined in terms of the solid angle between bins. %
In this way local peaks around a true source direction are removed, which is especially important near the poles where patches are more sparsely populated.

The 10 largest peaks in the smoothed histogram are initially picked as candidate source directions. %
Only peaks whose height are greater than $\peakSelectionTheshold$ times the height of the lowest peak (which is assumed to be due to noise) are retained.

\section{Implementation details}
\label{sec:implementation}
Specific parameter values for the implementation are given in Table~\ref{tab:parameter-values}. %

\begin{table}[h]
\begin{tabular}{ccc}%
\toprule
\bfseries Description [\si{\deg}] & \bfseries Value & \bfseries Units\\
\midrule
\ac{STFT} frame duration  & 4 & \si{\ms}\\
\ac{STFT} overlap factor  & 75 & \si{\percent}\\
\ac{SH} order of analsysis & 3 & ---\\
Covariance matrix time span  & 16 & \si{\ms}\\
Covariance matrix frequency span   &  350 & \si{\Hz}\\
Minimum frequency & 800 &  \si{\Hz}\\
Maximum frequency & 3500 &  \si{\Hz}\\
Histogram azimuth bin width & 2 & \si{\deg}\\
Histogram inclination bin width & 2 & \si{\deg}\\
Standard deviation of Gaussian smoothing kernel & 4 & \si{\deg}\\
Peak height ratio threshold, $\peakSelectionTheshold$  & 2 &  ---\\
\bottomrule
\end{tabular}
\caption{Parameter values used in implementation.}
\label{tab:parameter-values}
\end{table}

\section{Results}
\label{sec:results}

\subsection{Task 1}

Table~\ref{tab:task1} gives the accuracy of \ac{DOA} estimation using the \ac{SSPIV} method for Task 1, in which a single static source is present. %
Since it is known a priori that only one source is present, it is sufficient in the peak picking step to choose only the largest peak in the histogram. %
The average error in azimuth is \ang{8.6} while in elevation it is \ang{2.9} giving a combined solid angle error of \ang{9.2}.

\begin{table}[h]
\begin{tabular}{cccc}%
\toprule
\bfseries Rec \# & \bfseries Azimuth [\si{\deg}] & \bfseries Elevation [\si{\deg}] & \bfseries Combined [\si{\deg}]\\
\midrule
1   & 10.5  & 2.5 & 10.8\\
2   &  5.7  & 4.1 &  6.9\\
3   &  9.5  & 2.2 &  9.7\\
\addlinespace
Avg &  8.6  & 2.9 &  9.2\\
\bottomrule
\end{tabular}
\caption{Estimation error for source \acp{DOA} in Task 1.}
\label{tab:task1}
\end{table}

\subsection{Task 2}
\begin{figure}
  \centering
  \centerline{\includegraphics[width=\columnwidth]{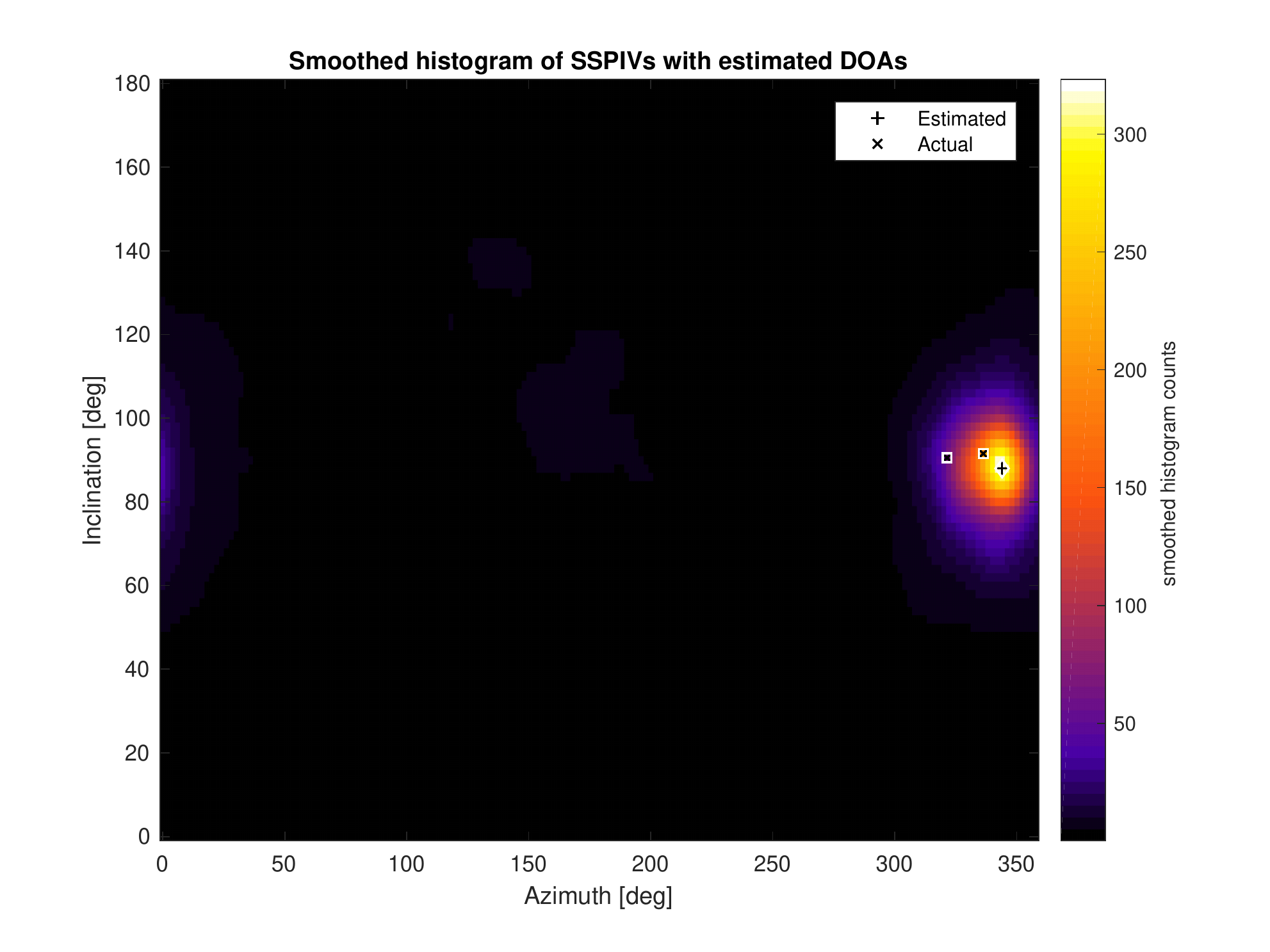}}
  \caption{Smoothed histogram of \acp{SSPIV} obtained for Task 2, recording 1 with ground truth and estimated \acp{DOA}.} 
  \label{fig:task2-rec1}
\end{figure}

\begin{figure}
  \centering     \includegraphics[width=\columnwidth]{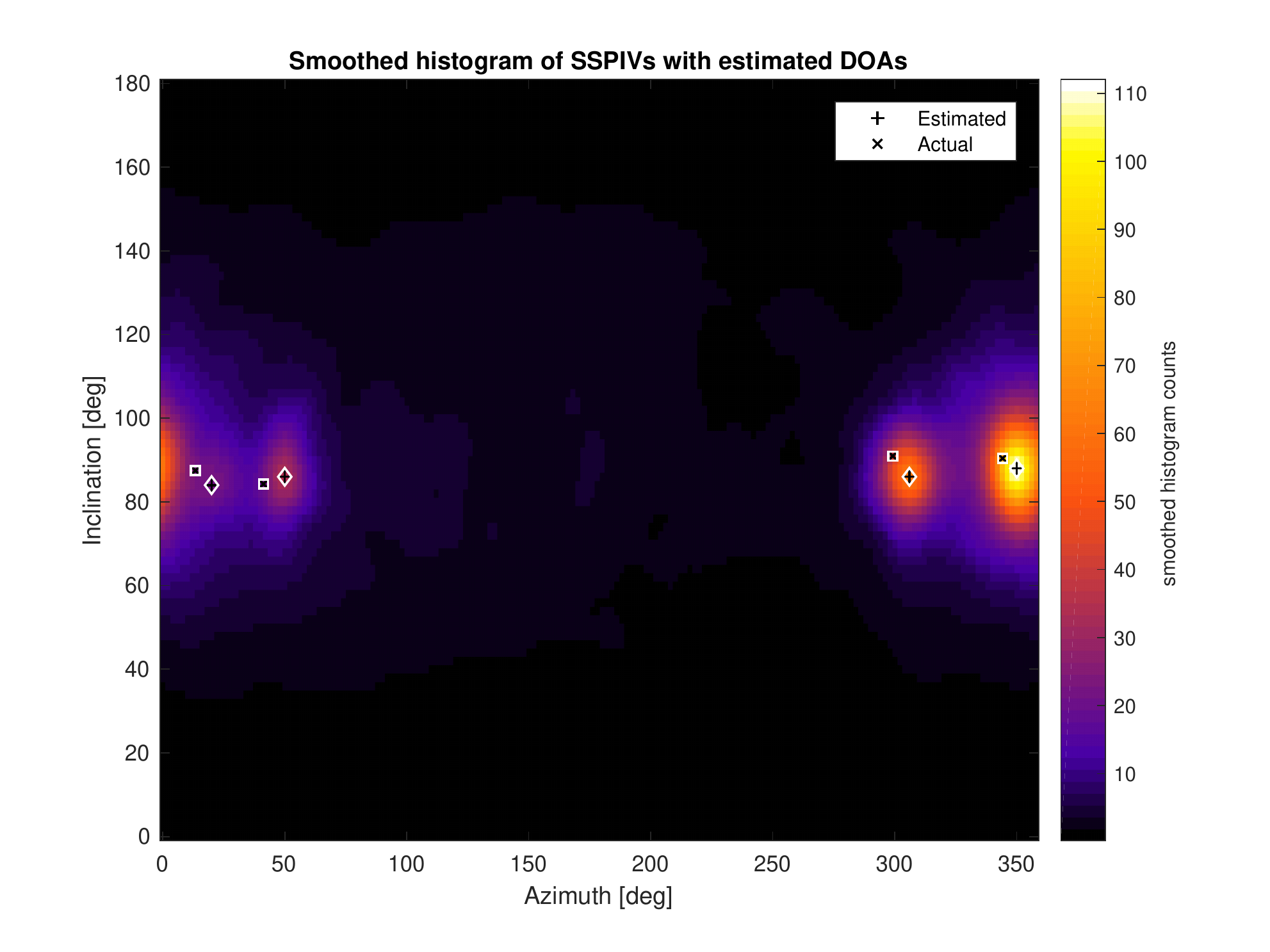}
  \caption{Smoothed histogram of \acp{SSPIV} obtained for Task 2, recording 2 with ground truth and estimated \acp{DOA}.} 
  \label{fig:task2-rec2}
\end{figure}

In Task 2 an unknown number of static sources are present. %
The task therefore includes both \ac{DOA} estimation and source counting. %
Table~\ref{tab:task2} gives the accuracy of \ac{DOA} estimation using the \ac{SSPIV} method. %
It can be seen that for recording 1 only one of the two sources are detected. %
Figure~\ref{fig:task2-rec1} shows the smoothed histogram of \acp{SSPIV} and there is clearly no second peak corresponding to the second source. %

In recordings 2 and 3, where more sources are active, all sources are identified. %
Figure~\ref{fig:task2-rec2} shows the smoothed histogram of \acp{SSPIV} for recording 2. %
The source at azimuth \ang{13} leads to a relatively small peak but is, nevertheless, detected. %

Neglecting the missed source, the average error in azimuth is \ang{8.9} while in elevation it is \ang{2.7} giving a combined solid angle error of \ang{9.5}. %
These values are remarkably close to those obtained for a single source, suggesting that the \ac{SSPIV} method is well suited to multiple source \ac{DOA} estimation in real-world environments.

\begin{table}[ht]
\begin{tabular}{ccccc}%
\toprule
\bfseries Rec \# & \bfseries Src \# & \bfseries Azimuth [\si{\deg}] & \bfseries Elevation [\si{\deg}] & \bfseries Combined [\si{\deg}]\\
\midrule
1 & 1 & 7.1 & 3.5 & 8.5\\
 & 2 &        --- &        --- &        ---\\
\addlinespace
2 & 1 & 8.8 & 1.7 & 8.9\\
 & 2 & 6.6 & 3.5 & 7.5\\
 & 3 & 5.8 & 2.4 & 6.3\\
 & 4 & 6.8 & 4.9 & 8.4\\
\addlinespace
3 & 1 & 19.2 & 0.9 & 19.2\\
 & 2 & 6 & 2.2 & 6.4\\
 & 3 & 10.7 & 2.6 & 10.9\\
\addlinespace
Avg & --- & 8.9 & 2.7 & 9.5\\
\bottomrule
\end{tabular}
\caption{Estimation error for source \acp{DOA} in Task 2.}
\label{tab:task2}
\end{table}

\section{Conclusions}
\label{sec:conclusions}
The application of the \ac{SSPIV} method to \ac{DOA} estimation for single and mulitple sources in the context of the \LOCATA\ Challenge has been described. %
Sources are estimated in both cases with a mean absolute error of about \ang{9}. 


\bibliographystyle{IEEEtran}
\bibliography{bib/sapstrings.bib,bib/sapref.bib,bib/local-bib.bib}

\end{sloppy}
\end{document}

%% file: mypreamble.tex
\pdfoutput=1
\usepackage{xcolor}
\usepackage{soul}
\usepackage{siunitx}
\DeclareSIUnit{\dbspl}{\dB~SPL}

\usepackage{overpic}
\usepackage{rotating}

\usepackage{booktabs}
\usepackage{csvsimple}

\usepackage{printlen} 

\usepackage{relsize}
\usepackage[printonlyused,withpage]{acronym} 

\usepackage{tikz}
\usetikzlibrary{shapes,arrows}

%% file: common_symbols.tex
\renewcommand{\deg}{\ensuremath{^{\circ}}}

\usepackage{amssymb}

%% file: symbol_definitions.tex
\def\LOCATA{\mbox{LOCATA}}

\newcommand{\peakSelectionTheshold}{\beta}

\newcommand{\stftFreq}{\nu}
\newcommand{\stftFrame}{\ell} 



%% file: myacronyms.txt
\acro{AIR}{acoustic impulse response}
\acro{DOA}{direction of arrival}
\acrodefplural{DOA}[DOAs]{directions of arrival}
\acro{DOF}{degree of freedom}
\acrodefplural{DOF}{degrees of freedom}
\acro{DTFT}{discrete time Fourier transform}
\acro{EWLS}{exponentially-weighted least squares}
\acro{GSC}{generalized sidelobe canceller}
\acro{IMU}{inertial measurement unit}
\acro{ISFT}{inverse \ac{SFT}}
\acro{LCMV}{linearly constrained minimum variance}
\acro{LTASS}{long-term average speech spectrum}
\acro{LTI}{linear time-invariant}
\acro{ML}{maximum likelihood}
\acro{MMSE}{minimum mean squared error}
\acro{MWF}{multichannel Wiener filter}
\acro{MVDR}{minimum variance distortionless response\acroextra{ beamformer}}
\acro{NCM}{noise covariance matrix}
\acro{NPD}{noise power distribution}
\acro{PSD}{power spectral density}
\acro{PWD}{plane-wave density}
\acro{PWDPSD}{\ac{PWD}\ac{PSD}}
\acro{RLS}{recursive least squares}
\acro{RMS}{root mean square}
\acro{RS}{recursive smoothing}
\acro{STFT}{short time Fourier transform}
\acro{SFT}{spherical Fourier transform}
\acro{SH}{spherical harmonic}
\acro{SNR}{signal-to-noise ratio}
\acro{TF}{time-frequency}
\acro{VAD}{voice activity detector}

\acro{SVD}{singular value decomposition}
\acro{PIV}{pseudointensity vector}
\acro{SSPIV}{subspace pseudointensity vector}
\acro{FS}{frequency smoothing}
\acro{TF}{time-freqeuncy}
\acro{WDO}{window-disjoint orthogonality}
\acro{DOA}{direction of arrival}
\acrodefplural{DOA}[DOAs]{directions of arrival}

%% file: block_diagram.tex
\newcommand{\STFT}{Short-time Fourier transform}
\newcommand{\SHT}{Spherical Fourier transform}
\newcommand{\MSC}{Mode strength compensation}
\newcommand{\SHTMSC}{Spherical Fourier transform with \\mode strength compensation}
\newcommand{\SSPIV}{Subspace pseudointensity vectors}
\newcommand{\SHIST}{Smoothed histogram}
\newcommand{\PEAKS}{Peak picking and pruning}

\newcommand{\micsig}{Microphone signals}
\newcommand{\estdoas}{Estimated \acp{DOA}}

\newcommand{\inlabel}{$\sigMicFreqVec(\stftFreq,\stftFrame)$}
\newcommand{\monoout}{$\sigBfOutFreqO(\stftFreq,\stftFrame)$}
\newcommand{\binoutl}{$\sigBfOutFreqL(\stftFreq,\stftFrame)$}
\newcommand{\binoutr}{$\sigBfOutFreqR(\stftFreq,\stftFrame)$}
\newcommand{\binenhl}{$\sigEnhOutFreqL(\stftFreq,\stftFrame)$}
\newcommand{\binenhr}{$\sigEnhOutFreqR(\stftFreq,\stftFrame)$}
\newcommand{\mask}{$\stftMask(\stftFreq,\stftFrame)$}
\newcommand{\pose}{$\rotCanonical(\stftFrame)$}
\newcommand{\worlddoa}{$\doaSourceWrtWorld(\stftFrame=0)$}
\newcommand{\monoBfLookDoa}{$\lookWrtArrayO(\stftFrame)=\doaSourceWrtArray(\stftFrame)$}
\newcommand{\binBfLookDoa}{$\lookWrtArrayL(\stftFrame)=\lookWrtArrayR(\stftFrame)=(90\deg,0\deg)$}

\newcommand{\BFTWO}{Bilateral \\ Beamformers}
\newcommand{\BFONE}{Reference \\ Beamformer}

\def\myyshift{0.2cm}
\def\mvd{2cm} 
\def\iod{1.5cm} 
\tikzstyle{block} = [draw, fill=white, rectangle, minimum height=1.2cm, minimum width=2cm]
\tikzstyle{sum} = [draw, fill=white, circle, node distance=1cm]
\tikzstyle{input} = [coordinate]
\tikzstyle{output} = [coordinate]
\tikzstyle{branchpoint} = [coordinate]
\tikzstyle{pinstyle} = [pin edge={to-,thin,black}]

\begin{tikzpicture}[auto,>=latex']
    \node [block, align=center] (STFT) {\STFT};
    \node [block, align=center, below of=STFT, node distance=\mvd] (SHT) {\SHTMSC};
    \node [block, align=center, below of=SHT, node distance=\mvd] (SSPIV) {\SSPIV};
    \node [block, align=center, below of=SSPIV, node distance=\mvd] (SHIST) {\SHIST};
    \node [block, align=center, below of=SHIST, node distance=\mvd] (PEAKS) {\PEAKS};
    
    \node [input, name=insig, above of=STFT, node distance=\iod] {};    
    \node [output, name=outest, below of=PEAKS, node distance=\iod] {};

    \draw [draw,->,thick] (insig) -- node {\micsig} (STFT.north);
    \draw [draw,->,thick] (STFT.south) -- node {} (SHT.north);   
    \draw [draw,->,thick] (SHT.south) -- node {} (SSPIV.north);     
    \draw [draw,->,thick] (SSPIV.south) -- node {} (SHIST.north);   
    \draw [draw,->,thick] (SHIST.south) -- node {} (PEAKS.north);   
    \draw [draw,->,thick] (PEAKS.south) -- node {\estdoas} (outest);   

\end{tikzpicture}